\documentclass[useAMS]{mn2e}

\usepackage{latexsym,graphicx}

\usepackage{color}

\newcommand\kms{{\rm\,km\,s^{-1}}}
\newcommand\msun{\rm\,M_\odot}

\newcommand\hii{H\,{\sc ii} \,}

\def\apgt{\ {\raise-.5ex\hbox{$\buildrel>\over\sim$}}\ }
\def\aplt{\ {\raise-.5ex\hbox{$\buildrel<\over\sim$}}\ }

\title[Field O stars: formed {\it in situ} or as runaways?]{Field O stars: formed {\it in situ} or as runaways?}
\author[V.V.Gvaramadze et al.]
       {V. V.~Gvaramadze,$^{1,2}$\thanks{E-mail: vgvaram@mx.iki.rssi.ru (VVG); cweidner@iac.es
       (CW); pavel@astro.uni-bonn.de (PK); jpflamm@astro.uni-bonn.de (JPA)}
        C.~Weidner,$^{3,4\star}$ P.~Kroupa,$^{5\star}$ and J.~Pflamm-Altenburg$^{5\star}$ \\
        $^{1}$Sternberg Astronomical Institute, Lomonosov Moscow State University, Universitetskij Pr. 13, Moscow 119992, Russia\\
        $^{2}$Isaac Newton Institute of Chile, Moscow Branch, Universitetskij Pr. 13, Moscow 119992, Russia\\
        $^{3}$Instituto de Astrof\'{i}sica de Canarias, C/ V\'{i}a L\'{a}ctea, s/n, E38205 La Laguna (Tenerife), Spain\\
        $^{4}$Scottish Universities Physics Alliance (SUPA), School of Physics and Astronomy, University of St. Andrews, North Haugh,\\
        St. Andrews, Fife KY16 9SS, UK\\
        $^{5}$Argelander-Institut f\"{u}r Astronomie, Universit\"{a}t Bonn, Auf dem H\"{u}gel 71, 53121 Bonn,
        Germany
        }
\begin{document}

\date{Accepted 2012 June 7.  Received 2012 June 7; in original form 2012 April 12}


\maketitle

\label{firstpage}

\begin{abstract}
A significant fraction of massive stars in the Milky Way and other
galaxies are located far from star clusters and star-forming
regions. It is known that some of these stars are runaways, i.e.
possess high space velocities (determined through the proper
motion and/or radial velocity measurements), and therefore most
likely were formed in embedded clusters and then ejected into the
field because of dynamical few-body interactions or
binary-supernova explosions. However, there exists a group of
field O stars whose runaway status is difficult to prove via
direct proper motion measurements (e.g. in the Magellanic Clouds)
or whose (measured) low space velocities and/or young ages appear
to be incompatible with their large separation from known star
clusters. The existence of this group led some authors to believe
that field O stars can form {\it in situ}. Since the question of
whether or not O stars can form in isolation is of crucial
importance for star formation theory, it is important to
thoroughly test candidates of such stars in order to improve
theory. In this paper, we examine the runaway status of the best
candidates for isolated formation of massive stars in the Milky
Way and the Magellanic Clouds by searching for bow shocks around
them, by using the new reduction of the {\it Hipparcos} data, and
by searching for stellar systems from which they could originate
within their lifetimes. We show that most of the known O stars
thought to have formed in isolation are instead very likely
runaways. We show also that the field {\it must contain} a
population of O stars whose low space velocities and/or young ages
are in apparent contradiction with the large separation of these
stars from their parent clusters and/or the ages of these
clusters. These stars (the descendants of runaway massive
binaries) cannot be traced back to their parent clusters and
therefore can be mistakenly considered as having formed {\it in
situ}. We argue also that some field O stars could be detected in
optical wavelengths only because they are runaways, while their
cousins residing in the deeply embedded parent clusters might
still remain totally obscured. The main conclusion of our study is
that there is no significant evidence whatsoever in support of the
{\it in situ} proposal on the origin of massive stars.
\end{abstract}


\section{Introduction}
\label{sec:intro}

One of the most important issues in the theory of star formation
is the still incomplete understanding of how massive ($\ga 10 \,
\msun$) stars form (Zinnecker \& Yorke 2007; McKee \& Ostriker
2007). At least four theories have been developed, the competitive
accretion scenario (Bonnell, Bate \& Zinnecker 1998; Bonnell, Vine
\& Bate 2004), collisional merging (Bonnell \& Bate 2002), the
single core collapse model (Krumholz \& McKee 2008)\footnote{Note
that models in which only massive stars form are idealised
analytical descriptions or gas-dynamical simulations with highly
unusual equations of state.}, and the fragmentation-induced
starvation model (Peters et al. 2010). It is therefore crucial to
find conclusive constraints for the formation of massive stars
from observations. One important piece of evidence can be deduced
from the formation sites of massive stars. While some theories
need other stars and gas around the massive stars and predict
their formation only within star clusters (competitive accretion,
collisional merging, fragmentation-induced starvation), the core
collapse model only needs a sufficiently massive and dense cloud
core and allows for an isolated origin of O stars, at the expense
of needing to postulate contrived initial conditions. Hence,
analysis of O stars located in isolation is necessary to deduce
whether or not they were formed {\it in situ} and thereby to
narrow down the existing theories.

Several studies (e.g. de Wit et al. 2004, 2005; Schilbach \&
R\"oser 2008) searched for O stars in apparent isolation and tried
to track down possible parent clusters. Although for most stars
possible parent clusters were found, these studies also resulted
in a number of candidate massive stars formed in isolation. These
candidates are often treated as O stars formed in genuine
isolation and thereby used to support one or another proposal on
the origin of massive stars. For example, Krumholz et al. (2010)
write ``de Wit et al. (2004, 2005) find that 4$\pm$ 2 per cent of
Galactic O stars formed outside of a cluster of significant mass,
which is consistent with the models presented here [...], but not
with the proposed cluster-stellar mass correlation". Similarly,
Selier, Heydari-Malayeri \& Gouliermis (2011) argue ``there is
[...] a statistically small percentage of massive stars
($\sim$5\%) that form in isolation (de Wit et al. 2005; Parker \&
Goodwin 2007)", and Franchetti et al. (2012) state that ``Among
the 227 Galactic O stars with $V<8$, $\sim 83 \, \%$ are in
clusters, $\sim 10\, \%$ are runaways, and only $5-10\, \%$ are
truly isolated (de Wit et al. 2004, 2005; Zinnecker \& Yorke
2007). Therefore, about $10\pm4$ \% of core-collapse SNRs are not
associated with other massive stars or star-forming regions
[...]".

The question whether O stars can form in isolation or not is
important well beyond the topic of star formation. If massive
stars need a clustered environment to form, star formation cannot
be a purely statistical process as it would result in a
non-trivial relation between the mass of a star cluster, $M_{\rm
cl}$, and the mass of the most-massive star, $m_{\rm max}$, formed
in this cluster (Weidner \& Kroupa 2006; Weidner, Kroupa \&
Bonnell 2010). As galaxies with a low star-formation rate form
only small star clusters (Weidner, Kroupa \& Larsen 2004), which
would not form any massive stars because of the $m_{\rm
max}-M_{\rm cl}$-relation, the integrated stellar populations
(integrated galactic stellar initial mass function, IGIMF) of such
galaxies could be quite different from the populations in
individual star clusters (see Weidner \& Kroupa 2005 and Kroupa et
al. 2011 for details). Thus, the problem of isolated massive star
formation remains highly relevant.

In this paper, we analyse if the known candidates for isolated
massive star formation in the Milky Way and the Magellanic Clouds
are actually unrecognised runaways and therefore were formed in
the clustered way. Before discussing individual objects, we review
the main mechanisms for the origin of runaway stars and methods
for their detection, and briefly discuss the possible origin of
low-velocity field O stars (Section\,\ref{sec:run}). In
Section\,\ref{sec:MW} we discuss candidates for isolated massive
star formation in the Milky Way, while those in the Magellanic
Clouds are discussed in Section\,\ref{sec:MC}. We summarize and
conclude in Section\,\ref{sec:sum}.

\section{Two subgroups of massive field stars}
\label{sec:run}

Massive stars that are not members of any known star cluster, OB
association or star-forming region are called field OB stars.
Observations show that about 20 per cent of Galactic O stars are
in the field (Gies 1987). It is also observed that massive stars
in other (nearby) galaxies (Magellanic Clouds, M33, etc) sometimes
lie outside star-forming regions (e.g. Madore 1978; Massey \&
Conti 1983; Kenyon \& Gallagher 1985). Although the O star census
in these galaxies is heavily incomplete, it is possible to
estimate the percentage of their massive field stars through the
distribution of Wolf-Rayet stars, the statistics of which is known
much better. The Wolf-Rayet stars are only slightly older than
their O-type progenitors, so that they should closely reflect the
distribution of massive stars in their parent
galaxies\footnote{Note that the Wolf-Rayet stars can be
accelerated to much higher speeds than the O stars (Gvaramadze,
Gualandris \& Portegies Zwart 2008). Correspondingly, some of them
can find themselves at much larger distances from their birth
clusters.}. The analysis of the distribution of Wolf-Rayet stars
in the Magellanic Clouds and in M33 by Massey et al. (1995) and
Neugent \& Massey (2011), respectively, showed that the percentage
of isolated Wolf-Rayet stars (and therefore that of isolated O
stars) in these galaxies is comparable to that of the Galactic
field O stars.

There are two subgroups of massive field stars: (i) OB stars with
high (say $> 30-40 \, \kms$; Blaauw 1961; Cruz-Gonz\'{a}lez et al.
1974) velocities (the so-called runaway stars; Blaauw 1961) and
(ii) low-velocity OB stars. About 20$-$30 per cent of the Galactic
field OB stars belong to the first subgroup (Blaauw 1961, 1993;
Gies 1987). The typical space velocity of stars in this subgroup
is several tens of $\kms$, although some of them possess much
higher velocities, up to several hundreds of $\kms$ (e.g.
Gvaramadze, Gualandris \& Portegies Zwart 2009 and references
therein).

It is believed that runaway stars are formed in star clusters and
then leave them because of two basic processes: (i) disruption of
a short-period binary system following the supernova explosion
(either symmetric or asymmetric) of one of the binary components
(the so-called binary-supernova scenario; Blaauw 1961; Stone 1991)
and (ii) dynamical three- or four-body encounters in dense stellar
systems (the so-called dynamical ejection scenario; Poveda, Ruiz
\& Allen 1967; Gies \& Bolton 1986). Obviously, some of the
runaway stars could form due to the combination of these two
processes, i.e. because of the dissolution of runaway massive
binaries (Pflamm-Altenburg \& Kroupa 2010; see also below).

The runaway stars can be revealed via several direct and indirect
methods. The direct methods are based on detection of high ($> 30
\, \kms$) peculiar transverse and/or radial velocities via proper
motion measurements (e.g. Blaauw 1961; Moffat et al. 1998;
Mdzinarishvili \& Chargeishvili 2005) and spectroscopy (e.g.
Massey et al. 2005; Evans et al. 2006, 2010), respectively. The
indirect indications of the runaway nature of some field OB stars
are the large (say $> 250$ pc) separation of these stars from the
Galactic plane (Blaauw 1961; van Oijen 1987) and the presence of
bow shocks around them (Gvaramadze \& Bomans 2008b; Gvaramadze et
al. 2011c). Revealing runaways via the detection of their
associated bow shocks is especially helpful for those of them
whose proper motions are still not available (e.g. in the
Magellanic Clouds) or are measured with a low significance. The
geometry of detected bow shocks can be used to infer the direction
of stellar motion and thereby to determine possible parent
clusters for the bow-shock-producing stars (Gvaramadze \& Bomans
2008a,b; Gvaramadze et al. 2010a, 2011b,c; Gvaramadze, Kroupa \&
Pflamm-Altenburg 2010b; Gvaramadze, Pfamm-Altenburg \& Kroupa
2011a).

There is, however, still no consensus on the origin of the
low-velocity subgroup of the massive field stars. A significant
fraction of these stars could be low-velocity runaways (e.g
Allison et al. 2010; Weidner, Bonnell \& Moeckel 2011); note that
the average escape velocity from the star cluster's potential well
is comparable to the average peculiar radial velocity of the field
O stars of $\sim 6.5 \kms$ (Gies 1987). Others could originate
because of rapid dissolution of star clusters following
residual-gas expulsion at the very beginning of cluster evolution
(e.g. Tutukov 1978; Kroupa, Aarseth \& Hurley 2001; Boily \&
Kroupa 2003a,b; Goodwin \& Bastian 2006; Weidner et al. 2007;
Baumgardt \& Kroupa 2007; Moeckel \& Bate 2010). Moreover, some
massive stars could be released into the field through the
dissolution of runaway massive binaries following the supernova
explosion of one of the binary companions (Pflamm-Altenburg \&
Kroupa 2010). The space velocity of the field stars produced in
this process is the vector sum of the ejection velocity of the
binary system, the orbital velocity of the star, and the kick
velocity imparted to the star by the stellar supernova remnant
(either neutron star or black hole) in the course of binary
disintegration (Tauris \& Takens 1998; Gvaramadze 2006, 2009). The
resulting velocity would be small if its components effectively
cancel each other.

\section{Galactic candidates for isolated massive star formation}
\label{sec:MW}

\begin{figure*}
 \resizebox{17.5cm}{!}{\includegraphics{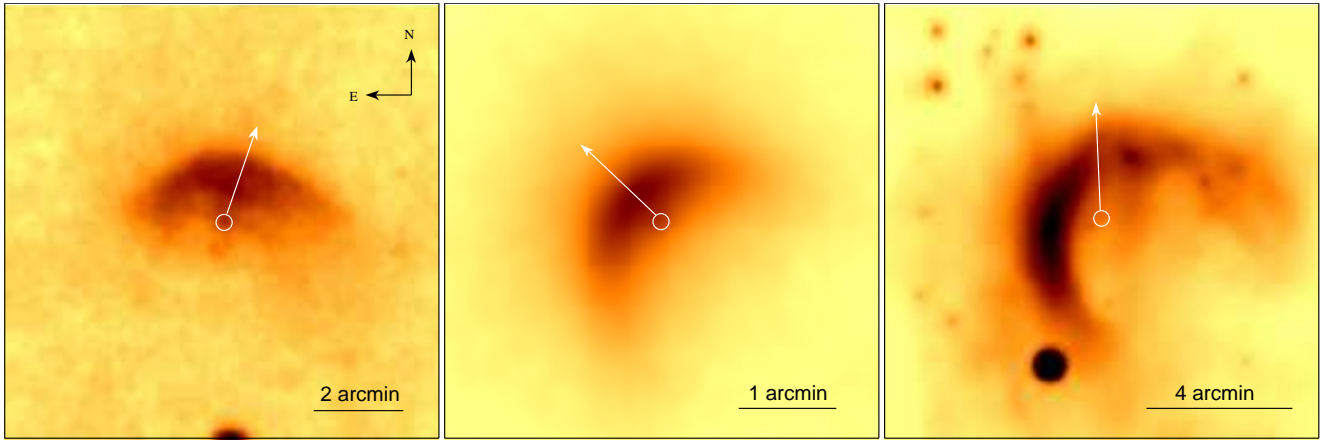}}
 \caption{{\it WISE} 22 $\mu$m images of bow shocks
associated with three field O stars: HD\,48279 (left panel),
HD\,57682 (middle panel) and HD\,153426 (right panel). The
positions of the stars are marked by circles. The directions of
the peculiar transverse velocities of the stars (derived from the
new reduction of the {\it Hipparcos} data) are indicated by
arrows. The orientation of the images is the same. See text for
details.} \label{fig:bows}
\end{figure*}

\subsection{de Wit et al. sample of O stars apparently formed in isolation}
\label{sec:deWit}

In the study of Galactic field O stars by de Wit et al. (2004,
2005), three possibilities for their origin were considered.
Namely, it was assumed that they are either (i) low-velocity
runaways, (ii) unrecognised runaways, or (iii) members of
unrecognised star clusters.

To check these possibilities, de Wit et al. used the catalogues of
Galactic O stars to select those which are not members of any
known cluster or OB association. It was found that $\approx 20$
per cent (43 out of 193) of O stars with $V< 8$ mag are located in
the field. For these stars de Wit et al. (2005) searched for known
young ($< 10$ Myr) star clusters or OB associations within the
projected distance of 65 pc (i.e. the drift distance which massive
stars wandering with a peculiar transverse velocity of $6.5 \,
\kms$ are likely to travel during their lifetimes). This search
resulted in the detection of possible parent clusters for seven
stars. Then, de Wit et al. (2005) excluded as field stars the
possible runaways using the {\it Hipparcos} data, the radial
velocity measurements and the distances from the Galactic plane.
To address the third possibility, de Wit et al. (2004) searched
for the presence of subparsec and tens of parsec scale clusters
around all 43 field stars using their own deep infrared imaging
and the Two-Micron All-Sky Survey (2MASS; Skrutskie et al. 2006),
respectively. They found stellar density enhancements near five
stars. One of these stars, HD\,57682, has a large peculiar
velocity (i.e. is a runaway star) so that the stellar density
enhancement around it was interpreted as a statistical noise
fluctuation (see also below).

Finally, de Wit et al. (2005) found that $\simeq 6$ per cent (11
out of 193) of O stars are not runaways and cannot be associated
with previously unrecognized clusters, and therefore were probably
formed {\it in situ}. Four of these eleven stars were regarded as
``the best examples for isolated Galactic high-mass star
formation" because of the presence of nearby indicators of recent
star formation (\hii regions, dark clouds, etc), so that the more
conservative estimate of the percentage of stars formed in
isolation is $4/193 \simeq 2$ per cent. Combining both estimates
in a single one, de Wit et al. (2005) concluded that 4$\pm$2 per
cent of Galactic O stars ``can be considered as formed outside a
cluster environment". This conclusion is often seen as `proof' for
the existence of isolated massive star formation or is used to
support one or another proposal related to the problem of massive
star formation (Parker \& Goodwin 2007; Camargo, Bonatto \& Bica
2010; Saurin, Bica \& Bonatto 2010; Krumholz et al. 2010; Lamb et
al. 2010; Franchetti et al. 2012).

\subsection{Narrowing down the de Wit et al. sample}
\label{sec:nar}

The percentage of O stars suggested by de Wit et al. (2005) to be
formed outside a cluster environment could be reduced two times
thanks to the study of Galactic field O stars by Schilbach \&
R\"{o}ser (2008). These authors retraced the orbits of 93 O stars
in the Galactic potential and found that six out of the eleven
stars from the de Wit et al. sample originate in known young star
clusters. One of these six stars, HD\,123056, belongs to the group
of the four stars considered by de Wit et al. (2005) as ``the best
examples for isolated Galactic high-mass star formation".
Moreover, Gvaramadze \& Bomans (2008b) demonstrated that one more
star from this group of four, HD\,165319, is a bow-shock-producing
(i.e. runaway) star, which most likely was ejected from the young
massive star cluster NGC\,6611. This latter discovery motivated us
to search for bow shocks around the remaining four of the eleven O
stars from the de Wit et al. sample that appear to be formed in
isolation, namely, HD\,48279, HD\,124314, HD\,193793, and
HD\,202124.

\begin{figure}
 \vspace*{-1.5cm}
 \resizebox{9cm}{!}{\includegraphics{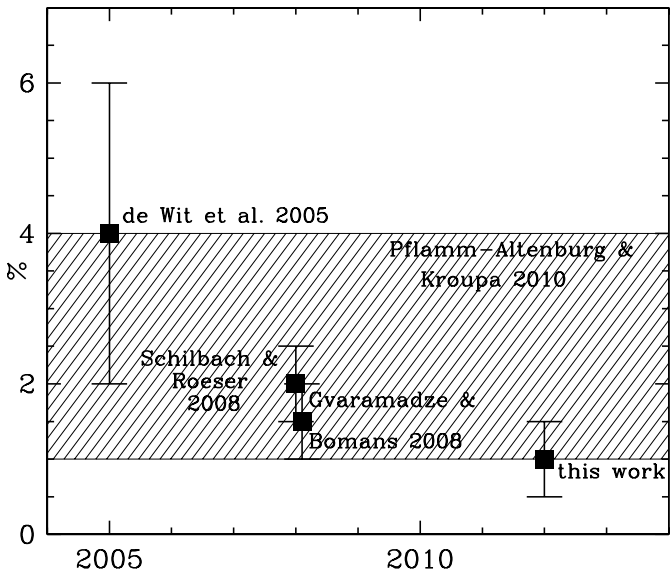}}
 \vspace*{-3cm}
 \caption{Evolution of the percentage of field O stars apparently
 formed in isolation with time. Data from de Wit et al. (2005), Schilbach
 \& R\"oser (2008), Gvaramadze \& Bomans (2008b), and this work. The
 current per cent of O stars apparently formed in isolation is fully
 consistent with what is expected (shaded area) from the two-step
 ejection mechanism for the origin of field O stars (Pflamm-Altenburg \& Kroupa
 2010). See text for details.}
 \label{fig:per}
\end{figure}

Our search for bow shocks was carried out by using the recently
released Mid-Infrared All Sky Survey carried out with the {\it
Wide-field Infrared Survey Explorer} ({\it WISE}; Wright et al.
2010). This survey provides images in four wavebands centred at
3.4, 4.6, 12 and 22\,$\mu$m (with angular resolution of 6.1, 6.4,
6.5 and 12.0 arcsec, respectively), of which the 22\,$\mu$m band
is most suitable for detection of bow shocks (e.g. Gvaramadze et
al. 2011c; Peri et al. 2012). Using the {\it WISE} data, we
discovered a bow shock generated by one more star, HD\,48279, from
the group of ``the best examples for isolated Galactic high-mass
star formation" (see the left panel of Fig.\,\ref{fig:bows} for
the 22\,$\mu$m image of this bow shock). This leaves us with three
O stars apparently formed outside a cluster environment, so that
the percentage of these stars is reduced to 1.0$\pm$0.5 per cent
(see Fig.~\ref{fig:per} for the evolution of this reduction).

This small percentage of isolated O stars can easily be understood
if one takes into account the fact that some massive stars can
find themselves in the field because of the combined effect of
dynamical ejection of massive binaries and their subsequent
disruption following the supernova explosion of one of the binary
components. The vast majority of field stars resulting from this
{\it two-step ejection process} cannot be traced back to their
parent clusters and therefore can be mistakenly considered as
formed {\it in situ} (Pflamm-Altenburg \& Kroupa 2010). It is
important to note that the supernova explosion in a runaway binary
can not only re-direct or accelerate the companion star, but can
also decelerate or even effectively stop it, so that the peculiar
space velocity of the newly formed single field star could be much
smaller than the ejection velocity of the runaway binary (Tauris
\& Takens 1998; Gvaramadze 2006, 2009). The obvious consequence of
this effect is that the field {\it must contain} O stars, whose
low space velocities are in apparent contradiction with the large
separation of these stars from their parent clusters.

\begin{table*}
\caption{Proper-motion measurements and heliocentric radial
velocities (when available) for three O stars apparently formed in
isolation (first three rows) and for three bow-shock-producing
field O stars. For each star, the components of the peculiar
transverse velocity (in Galactic coordinates), the peculiar radial
velocity, and the total space velocity is calculated and added to
the table.} \label{tab:prop}
 \begin{minipage}{\textwidth}
\begin{tabular}{cccccccc}
\hline \hline Star &  $\mu _\alpha \cos \delta ^{a}$ & $\mu
_\delta ^{a}$ &
$v_{\rm r,hel}$ & $v_{\rm l}$ & $v_{\rm b}$ & $v_{\rm r}$ & $v_{\rm tot}$ \\
~ & mas ${\rm yr}^{-1}$ & mas ${\rm yr}^{-1}$ & $\kms$ & $\kms$ & $\kms$ & $\kms$ & $\kms$\\
\hline
HD\,124314 & $-3.85\pm0.87$ & $-1.98\pm0.69$ & -- & $14.6\pm4.2$ & $4.2\pm3.5$ & -- & $\geq 15.2\pm4.1$ \\
HD\,193793 & $-5.20\pm0.37$ & $-1.63\pm0.33$ & -- & $7.2\pm2.7$ & $33.3\pm2.8$ & -- & $\geq 34.1\pm2.8$ \\
HD\,202124 & $-1.30\pm0.57$ & $-5.99\pm0.45$ & $-23.6\pm3.3 ^b$ & $0.9\pm7.7$ & $-39.4\pm7.8$ & $2.1\pm3.3$ & $-39.5\pm7.8$ \\
\hline
HD\,48279 & $-1.86\pm0.83$ & $2.73\pm0.72$ & $15.0\pm5.0^c$ & $-20.5\pm5.8$ & $3.4\pm6.3$ & $-19.6\pm5$ & $28.6\pm5.4$ \\
HD\,57682 & $10.46\pm0.45$ & $13.38\pm0.34$ & $23.0\pm2.0^c$ & $-23.2\pm1.9$ & $89.6\pm2.2$ & $-9.9\pm2.0$ & $93.1\pm2.2$ \\
HD\,153426 & $-0.58\pm0.98$ & $0.53\pm0.54$ & $-6.4\pm5.0^c$ & $19.2\pm6.8$ & $13.5\pm7.7$ & $18.3\pm5.0$ & $29.8\pm6.4$ \\
\hline
\end{tabular}
\end{minipage}
$^{a}$van Leeuwen (2007); $^{b}$Kharchenko et al. (2007);
$^{c}$Evans 1967.
\end{table*}
\begin{table*}
  \caption{Details of three O stars apparently formed in isolation (first three rows)
  and of three bow-shock-producing field O stars.}
  \label{tab:phot}
  \begin{minipage}{\textwidth}
\begin{tabular}{ccccccccc}
\hline \hline
Star & Spectral & $B^f$ & $V^f$ & $J^g$ & $K_{\rm s} ^g$ & $A_V$ & $A_{K_{\rm s}}$ & Distance \\
 & type & mag & mag & mag & mag & mag & mag & kpc \\
\hline
HD\,124314 & O6\,V(n)((f))$^a$ & 6.85 & 6.64 & 6.18 & 6.09 & 1.49 & 0.20 & 1.04 \\
HD\,193793 & WC7+O5.5$^b$ & -- & -- & -- & -- & -- & -- & 1.67$^{h}$ \\
HD\,202124 & O9.5\,Iab$^c$ & 8.06 & 7.82 & 7.18 & 7.09 & 1.54 & 0.20 & 3.20 \\
\hline
HD\,48279 & O8\,V$^d$ & 8.04 & 7.89 & 7.65 & 7.69 & 1.30 & 0.11 & 1.64 \\
HD\,57682 & O9\,V$^e$ & 6.23 & 6.42 & 6.81 & 6.94 & 0.25 & 0.05 & 1.11 \\
HD\,153426 & O9\,II-III$^a$ & 7.61 & 7.47 & 7.06 & 7.01 & 1.24 & 0.17 & 1.94 \\
\hline
\end{tabular}
\end{minipage}
$^a$Walborn (1973); $^b$Fahed et al. (2011); $^c$Walborn (1971);
$^d$Walborn (1970); $^e$Johnson \& Morgan (1953); $^f$Mermilliod
(1991); $^g$Cutri et al. (2003); $^h$Monnier et al. (2011).
\end{table*}

Moreover, some of the field O stars could be the products of the
merging of the components of runaway binary systems. Imagine a
tight binary (composed of two $10 \, \msun$ stars) ejected from a
cluster with a velocity of $30 \, \kms$. During the main sequence
stage the binary components increase their radii several times so
that the binary could merge into a single O star if the system was
sufficiently tight\footnote{Note that in the course of dynamical
ejection binary systems acquire high eccentricities (typically
$\ga 0.6$; Hills 1975; Hoffer 1983), which also facilitates the
merging of binary components.}. If the binary merged after
$\approx 20$ Myr since the ejection event (i.e. at a distance of
$\approx 500$ pc from the birth place), the resulting rejuvenated
star (blue straggler) would appear much younger than its parent
cluster, while the cluster itself could be completely dissolved by
that time.

The above considerations can explain the origin of single field O
stars (unless the ejected systems were triple or of higher
multiplicity; cf. Gvaramadze \& Menten 2012). It turns out,
however, that one of the remaining three O stars apparently formed
in isolation, HD\,124314, is a candidate single-line spectroscopic
binary (Feast, Thackeray \& Wesselink 1955), while other one,
HD\,193793, is a massive binary composed of a WC7 and an O5.5 star
(Fahed et al. 2011). Let us examine the possibility that both
these systems are runaways, i.e. were dynamically ejected from
putative parent clusters.

To check the possible runaway status of these two stars
(HD\,124314, HD\,193793) and the third star (HD\,202124)
apparently formed in isolation, we determine their space
velocities using proper motion measurements from the new reduction
of the {\it Hipparcos} data by van Leeuwen (2007). These
measurements along with the heliocentric radial velocity of
HD\,202124 (taken from Kharchenko et al. 2007) are summarized in
Table\,\ref{tab:prop}. To convert the observed proper motions and
the radial velocity into the peculiar transverse and radial
velocities of the stars, we used the Galactic constants $R_0 =
8.0$ kpc and $\Theta _0 =240 \, \kms$ (Reid et al. 2009) and the
solar peculiar motion
$(U_{\odot},V_{\odot},W_{\odot})=(11.1,12.2,7.3) \, \kms$
(Sch\"onrich, Binney \& Dehnen 2010). The distances to HD\,124314
and HD\,202124 were determined using their $B$ and $V$ magnitudes
from the Catalogue of Homogeneous Means in the $UBV$ System by
Mermilliod (1991), the $J$ and $K_{\rm s}$ magnitudes from 2MASS
(Cutri et al. 2003) (see Table\,\ref{tab:phot} for a summary of
these magnitudes), and the photometric calibration of optical and
infrared magnitudes for Galactic O stars by Martins \& Plez
(2006). For HD\,193793 we used the distance derived by Monnier et
al. (2011) through spectroscopic and interferometric measurements
of the orbit of this binary system.

The visual and $K_{\rm s}$-band extinctions towards the stars
(given in columns 7 and 8 of Table\,\ref{tab:phot}) were
calculated using the relationships:
\begin{equation}
A_V \, = \, 3.1E(B-V) \, , \label{eqn:AV}
\end{equation}
\begin{equation}
A_{K_{\rm s}} = 0.66E(J-K_{\rm s}) \, , \label{eqn:AK}
\end{equation}
where we adopted the extinction law from Rieke \& Lebofsky (1985)
and the standard total-to-selective absorption ratio $R_V =3.1$.
The mean distances obtained from the optical and the infrared
photometry are given in column 9 of Table\,\ref{tab:phot}.

The derived components of the peculiar transverse velocity in the
Galactic coordinate system and the peculiar radial velocity of
HD\,202124 are listed in columns 5, 6 and 7 of
Table\,\ref{tab:prop}. For the error calculation, only the errors
of the proper motion and the radial velocity measurements were
considered.

It follows from Table\,\ref{tab:prop} that HD\,193793 and
HD\,202124 are running away from the Galactic plane and that their
space velocities (see column 8 of Table\,\ref{tab:prop}) are $>30
\, \kms$, so that both stars are runaways in the classical sense.
It should be noted here that the non-detection of bow shocks
around these stars does not contradict their runaway status. The
point is that only a small fraction ($\approx 20$ per cent) of
runaway OB stars produce (observable) bow shocks (van Buren,
Noriega-Crespo, Dgani 1995). The most reliable explanation of this
empirical fact is that the majority of runaway stars are moving
through a low density, hot medium, so that the emission measure of
their bow shocks is below the detection limit or the bow shocks
cannot be formed at all because the sound speed in the local
interstellar medium is higher than the stellar space velocity
(e.g. Huthoff \& Kaper 2002).

Table\,\ref{tab:prop} also shows that the transverse peculiar
velocity of the candidate single-line spectroscopic binary
HD\,124314 is well below $30 \, \kms$. Since there are no reliable
radial velocity measurements for HD\,124314, one cannot exclude
the possibility that this star is a runaway as well. On the other
hand, HD\,124314 is an O6\,V(n)((f)) star, where `(n)' refers to
its fast ($v\sin i \simeq 250 \, \kms$; Penni 1996) rotation,
which could be caused by the mass transfer in the close binary
system. The natural consequence of the mass transfer is that the
mass receiver not only spins up but can also be rejuvenated (e.g.
Dray \& Tout 2007), so that it would appear much younger than its
actual age. If the rejuvenated star (blue straggler) is a member
of a runaway binary then the actual distance travelled by the
system could be much larger than that inferred from the apparent
age of the blue straggler and the peculiar transverse velocity of
the system derived from proper motion measurements. Moreover, the
absence of the secondary star contribution to the spectrum of
HD\,124314 might imply that the secondary is either a neutron star
or a black hole (i.e. HD\,124314 might be a post-supernova binary
system). It is therefore possible that the space velocity of the
binary was reduced (and re-oriented) due to the kicks caused by
the mass loss from the system and the asymmetry of the supernova
explosion (Stone 1982). In this case, the actual separation of the
binary from the parent cluster could also be much larger than
follows from the current (transverse) peculiar velocity and the
apparent age of the rejuvenated star.

For the sake of completeness, we note that the use of the {\it
WISE} data led to the discovery of bow shocks generated by two
additional field O stars, HD\,57682 and HD\,153426 (see the middle
and the right panels of Fig.\,\ref{fig:bows}), which, according to
de Wit et al. (2004), are surrounded by stellar density
enhancements. This discovery confirms the already known runaway
status of HD\,57682 (see Comer\'{o}n, Torra \& Gomez 1998) and
implies that HD\,153426 is a runaway as well. Moreover, one more
field O star (HD\,195592) associated with a stellar density
enhancement is a known runaway and bow-shock-producing star
(Noriega-Crespo, van Buren \& Dgani 1997; Peri et al. 2012), so
that at least three of the five stellar density enhancements
detected by de Wit et al. (2004) around field O stars are noise
fluctuations or chance superpositions.

Using the same procedure as for the three stars apparently formed
in isolation, we calculated the distances to and the peculiar
velocities of the three bow-shock-producing stars shown in
Fig.\,\ref{fig:bows} (see Tables\,\ref{tab:prop} and
\ref{tab:phot}). As expected, all three stars have space
velocities large enough to classify them as runaways, while the
orientation of their peculiar transverse velocities agree fairly
well with the orientation of the symmetry axis of the bow shocks
(see Fig.\,\ref{fig:bows}).

\section{Candidates for isolated massive star formation in the Magellanic Clouds}
\label{sec:MC}

The problem of isolated massive star formation was also widely
discussed with regard to the Magellanic Clouds (Massey et al.
1995; Massey 1998; Oey, King \& Parker 2004).

\subsection{Very massive field stars in the Large Magellanic Cloud}

The study of the massive star population in the Large Magellanic
Cloud (LMC) by Massey et al. (1995) has shown that several very
massive (O3$-$O4-type) stars are located at $\approx 100-200$ pc
(in projection) from known star clusters and OB
associations\footnote{In their footnote 5, Massey et al. (1995)
list eight such stars; some of them were later re-classified as
O2-type ones (Massey et al. 2005).}. This finding was interpreted
as indicating that the field can produce stars as massive as those
born in clusters (Massey et al. 1995; Massey 1998). In their
reasoning, Massey et al. (1995) proceed from the general belief
that most runaways originate in the course of disruption of
massive tight binaries following the supernova explosion of one of
the binary components (Blaauw 1961). From this it follows that an
early-type massive star (such as an O3 or O4) would simply have no
time to travel far from the birth cluster due to the youthfulness
of this phase (Massey et al. 1995). Today it is known, however,
that a more efficient channel for producing massive runaways is
based on dynamical few-body encounters in the dense cores of
embedded clusters (Poveda et al. 1967; Leonard \& Duncan 1990;
Clarke \& Pringle 1992; Pflamm-Altenburg \& Kroupa 2006;
Gvaramadze \& Gualandirs 2011; Fujii \& Portegies Zwart 2011;
Banerjee, Kroupa \& Oh 2012). In contrast to the binary-supernova
ejection mechanism, the gravitational slingshot effect starts to
produce runaways already in the course of cluster formation or at
the very beginning of cluster dynamical evolution.

The large separations from the possible parent clusters and the
young ($\approx 2$ Myr) ages of the very massive field stars imply
that their (transverse) peculiar velocities should be as high as
$\approx 50-100 \, \kms$ (Walborn et al. 2002)\footnote{The large
offsets from the parent clusters and the high peculiar velocities
are not unusual for Galactic very massive field stars as well. For
example, the bow-shock producing O4\,If star BD+43$\degr$ 3654
ejected from the Cyg\,OB2 association is located at $\approx 80$
pc from the core of the association (Comer\'{o}n \& Pasquali 2007;
Gvaramadze \& Bomans 2008a), while its space velocity is $\approx
70 \, \kms$ (Gvaramadze \& Gualandris 2011). Other good examples
of Galactic very massive field stars are two O2\,If*/WN6 stars,
which are located at $\approx 40-60$ pc in projection from their
likely parent cluster Westerlund\,2 (Roman-Lopes et al. 2011).},
provided that these stars escaped into the field soon after the
cluster formation. The runaway interpretation of the very massive
field stars in the LMC received strong support after the discovery
that some of them (including two O2-type stars, Sk$-$67$\degr$22
and BI\,237, listed in Massey et al. 1995) indeed have such very
high ($\approx 100-150 \, \kms$) peculiar (radial) velocities
(Massey et al. 2005; Evans et al. 2010). Further support to the
runaway interpretation of these stars comes from the detection of
a bow shock around BI\,237, whose orientation suggests that this
O2\,V((f*)) star (Massey et al. 2005) was ejected from the
association LH\,82 (Gvaramadze et al. 2010b; see also below).

There are three important issues that need to be kept in mind when
assessing the likelihood of association between very young
isolated massive stars and nearby stellar systems.

First, massive field stars might be blue stragglers dynamically
ejected from their birth clusters. Some of them could already be
formed in the parent clusters via merging of less massive stars in
the course of close binary-binary encounters. The numerical
experiments by Leonard (1995) showed that a significant fraction
of such merger products is ejected into the field with peculiar
velocities large enough to be classified as runaways. Moreover,
dynamical encounters can also produce runaway binaries (e.g.
Leonard \& Duncan 1990; Kroupa 1998; Oh \& Kroupa 2012), which
then can produce blue stragglers by mass transfer or merging
caused by stellar evolution (e.g. Gvaramadze \& Bomans 2008a).
Observations show that most O stars (both in clusters/associations
and in the field) are binaries or higher-order multiples (e.g.
Chini et al. 2012 and references therein). More importantly,
$\approx 70\pm10$ per cent of runaway O stars are binaries as well
(Chini et al. 2012)\footnote{This observation clearly show that
dynamical few-body encounters are the most important channel for
production of runaways.}. This fact along with the high proportion
of massive binaries with short periods (Mermilliod \& Garc\'{i}a
2001) and unit mass ratio (Pinsonneault \& Stanek 2006; Kobulnicky
\& Fryer 2007) makes the rejuvenation process in runaway binaries
common. From this it follows that the actual distances travelled
by rejuvenated massive field stars might be much greater than
those inferred from the apparent young ages of these stars and the
assumed (plausible) ejection velocities.

\begin{figure}
\includegraphics[width=8.5cm]{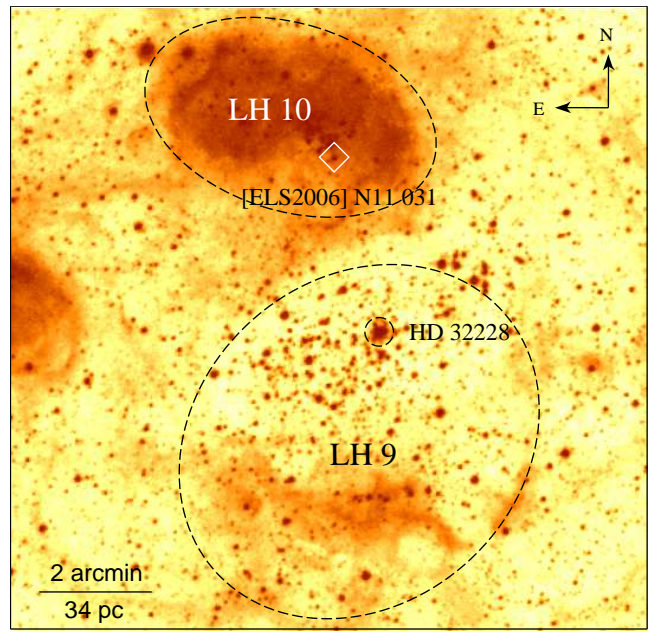}
\centering \caption{DSS-II red band image of the star-forming
region N11 with the approximate boundaries of two rich OB
associations, LH\,9 and LH\,10, indicated by dashed ellipses. The
compact massive star cluster HD\,32228 within the association
LH\,9 is indicated by a dashed circle. The position of the
ON2\,III(f*) star [ELS2006]\,N11\,031 is marked by a diamond. See
text for details.} \label{fig:LH9}
\end{figure}

We speculate that the ON2\,III(f*) (Walborn et al. 2004) star
[ELS2006]\,N11\,031 (currently located within the confines of the
OB association LH\,10) might be such a runaway blue straggler. In
Fig.\,\ref{fig:LH9} we show the Digitized Sky Survey II (DSS-II)
red band (McLean et al. 2000) image of the star-forming region N11
(Henize 1956), which is the second largest H\,{\sc ii} region in
the LMC after 30\,Doradus. N11 is composed of several rich OB
associations, two of which, LH\,9 and LH\,10, are outlined in
Fig.\,\ref{fig:LH9} by dashed ellipses\footnote{The approximate
boundaries of these associations are taken from Bica et al.
(1999).}. The radial velocity of [ELS2006]\,N11\,031 is $\approx
30 \, \kms$ greater than the median radial velocity of stars in
N11 (Evans et al. 2006), which implies that this star might be a
runaway. Moreover, [ELS2006]\,N11\,031 is surrounded by a bow
shock-like structure (see fig.\,7a in Gvaramadze et al. 2010b),
whose orientation suggests that [ELS2006]\,N11\,031 is running
away from the $\approx 3.5$ Myr old (Walborn et al. 1999) massive
compact cluster HD\,32228\footnote{Until relatively recently,
HD\,32228 was believed to be one of the brightest single stars in
the LMC.}, which is located at $\approx 46$ pc to the south of the
star (see Fig.\,\ref{fig:LH9}). If [ELS2006]\,N11\,031 was indeed
ejected from HD\,32228, then it should be a blue straggler because
its (apparent) age is about two times younger than that of the
cluster. Correspondingly, the transverse peculiar velocity of
[ELS2006]\,N11\,031 should be $\geq 12 \, \kms$.

Secondly, the presence of very massive stars in clusters and
associations does not necessary mean that these stellar systems
are as young as their most massive members (Gvaramadze \& Bomans
2008a,b; Gvaramadze et al. 2011c). Some of these stars could be
the merger products of dynamical few-body encounters (e.g.
Portegies Zwart et al. 1999; Walborn et al. 1999), while others
could be rejuvenated in the course of close binary evolution. Both
processes can produce a wide range of apparent ages and can
therefore explain the age spread often observed (e.g. Massey 2011)
in star clusters and associations. Moreover, some relatively old
OB associations could be `rejuvenated' by young massive stars
injected into these associations from nearby stellar systems
(Gvaramadze \& Bomans 2008a; Gvaramadze et al. 2011a).

\begin{figure*}
\includegraphics[width=17cm]{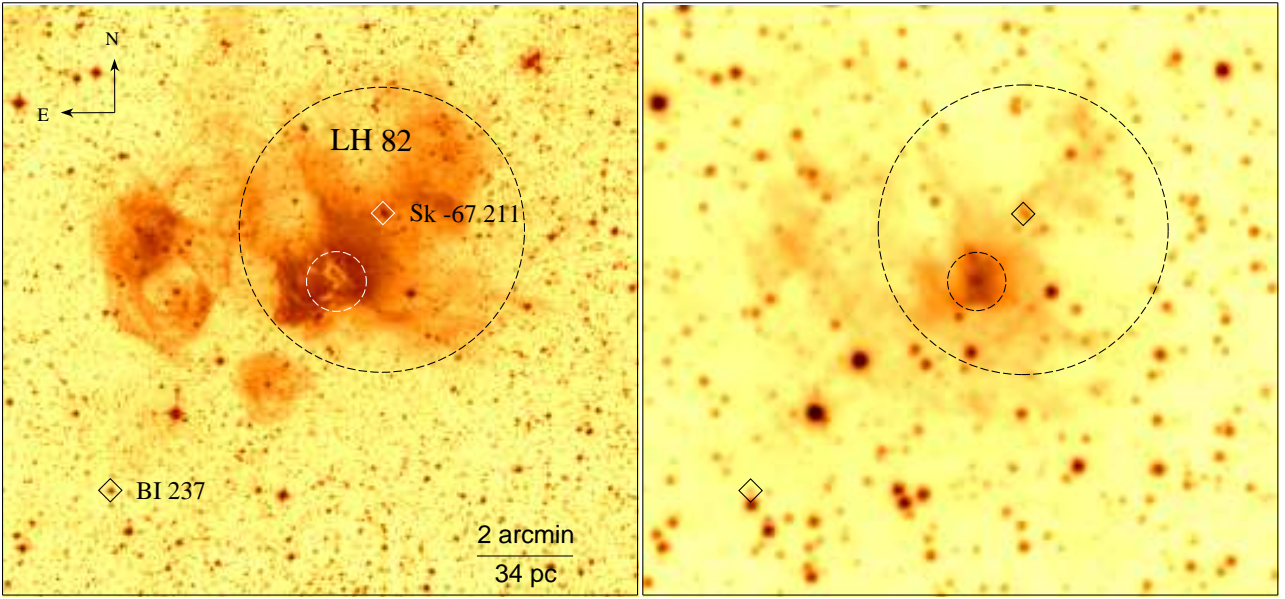}
\centering \caption{{\it Left}: DSS-II red band image of the field
containing the association LH\,82 (indicated by a large dashed
circle) and two O2-type stars, BI\,237 and Sk$-$67$\degr$211
(marked by diamonds). The position of a putative embedded star
cluster is shown by a small dashed circle (see text for details).
{\it Right}: {\it WISE} 3.4 $\mu$m image of the same field.}
\label{fig:BI237}
\end{figure*}

Thirdly, it is well known that stars in the Milky Way typically
form in dense embedded clusters (Lada \& Lada 2003) with a
characteristic radius of $\la 1$ pc, which is independent of
cluster mass (Kroupa \& Boily 2002; Marks \& Kroupa 2012). If star
formation in other galaxies follows the same physics as it does in
the Galaxy, then the known LMC clusters and OB associations should
have been born from configurations that are as dense as the
embedded clusters in the Galaxy. A case in point is R136 which,
like HD\,32228, was, for a long time, thought to be a single very
massive star, but is today known to be a compact very young
massive star cluster (e.g. Weigelt \& Baier 1985; Massey \& Hunter
1998). Catching an embedded cluster which has a diameter smaller
than a pc in the LMC would be difficult due to the high
obscuration and compactness of such an object. Thus, it may even
be that some of the isolated very massive stars may have been
ejected from a compact embedded cluster which remains
undiscovered. In other words, some massive stars can be detected
in optical wavelengths only because they are runaways, while their
cousins residing in the deeply embedded parent clusters might
still remain totally obscured (see Gvaramadze et al. 2010a for
possible Galactic examples of such runaways).

As mentioned above, the orientation of the bow shock around the
O2\,V((f*)) star BI\,237 suggests that this star is running away
from the association LH\,82. To eject dynamically such a massive
star, the association should initially contain a dense core of
massive stars, which would currently be spread over the
association. However, using the SIMBAD and the VizieR data bases
we found only two known O-type stars within the confines of
LH\,82, of which the O2\,III(f*) star (Walborn et al. 2004)
Sk$-$67$\degr$211 is comparable in mass with BI\,237. In
Fig.\,\ref{fig:BI237} we present the DSS-II red band and the {\it
WISE} 3.4 $\mu$m images of the field containing LH\,82 and the two
O2-type stars. One can see that these stars are located on the
opposite sides of a dark lane (marked in the left panel of
Fig.\,\ref{fig:BI237} by a small dashed circle), which coincides
with a compact infrared nebula in the {\it WISE} image (see the
right panel of Fig.\,\ref{fig:BI237}). We propose that this lane
might represent the parental cloud of a deeply embedded young
massive cluster from which the two O2-type stars were dynamically
ejected. If our proposal is correct, then the peculiar transverse
velocities of BI\,237 and Sk$-$67$\degr$211 should be $\approx 50$
and $12 \, \kms$, respectively, provided that both stars were
ejected $\approx 2$ Myr ago. Using these estimates and the
peculiar radial velocity of BI\,237 of $120 \, \kms$ (Massey et
al. 2005), one finds that the total space velocity of this star is
$\approx 130 \, \kms$ (cf. Gvaramadze et al. 2010b).

Recent numerical scattering experiments by Gvaramadze \&
Gualandris (2011) showed that three-body dynamical encounters
between a massive binary and a single massive star can easily
produce massive 
runaways with space velocities of $100-150 \, \kms$. Moreover,
$N$-body simulations of initially fully mass-segregated and
binary-rich massive star clusters by Banerjee et al. (2012)
clearly demonstrated that massive runaways represent the most
probable type of runaways produced by such clusters (see also
Section\,\ref{sec:other}). These results provide a natural
explanation of the origin of very massive field stars both in the
Magellanic Clouds and in the Milky Way.

Although it was realised that most (if not all) isolated massive
stars in the Magellanic Clouds could be runaways (Walborn et al.
2002, 2011; Foellmi, Moffat \& Guerrero 2003; Massey et al. 2005;
Evans et al. 2006, 2010; Brandl et al. 2007; Gvaramadze et al.
2010b, 2011a; Gvaramadze \& Gualandris 2011; Fujii \& Portegies
Zwart 2011; Banerjee et al. 2012), the possibility of massive star
formation outside of a cluster environment or in low-mass, sparse
clusters remains discussed. Recently, several instances of
isolated massive star formation in the Magellanic Clouds have been
proposed in the literature (Lamb et al. 2010; Selier et al. 2011;
Bestenlehner et al. 2011). Let us discuss them by turn.

\subsection{Lamb et al. sample of O stars formed in isolation}
\label{sec:lamb}

\begin{table}
\caption{Eight isolated OB stars in the SMC (Lamb et al. 2010).}
\label{tab:lam}
\begin{minipage}{\textwidth}
\begin{tabular}{cccc}
\hline \hline
Star & Spectral & $v_{\rm r} ^{a}$ & $v_{\rm r}$ \\
& type & $\kms$ & $\kms$ \\
\hline
[MLD95] SMC\,16 & O9\,V$^{a}$ & 121$\pm21$ & 167$\pm$11$^{d}$ \\
AzV\,58 & B0.5\,III$^{a}$ & 146$\pm$11 & -- \\
AzV\,67 & O8\,V$^{a}$ & 159$\pm13$ & 179$\pm$11$^{d}$ \\
AzV\,106 & B1\,II$^{a}$ & 150$\pm$12 & -- \\
AzV\,186 & O8\,III((f))$^{b}$ & 159$\pm$10 & 189$\pm$7$^{d}$ \\
AzV\,223 & O9.5\,II$^{c}$ & 189$\pm$7 & 190$^{c}$ \\
AzV\,226 & O7\,IIIn((f))$^{b}$ & 146$\pm$21 & 208:$^{e}$ \\
AzV\,302 & O8.5\,V$^{a}$ & 161$\pm$11 & 140$\pm$9$^{d}$ \\
\hline
\end{tabular}
\end{minipage}
 $^{a}$Lamb et al. (2010); $^{b}$Evans et al. (2004); $^{c}$Massey
 et al. (2009); $^{d}$Evans \& Howarth (2008); $^{e}$Evans et al. (2006).
\end{table}

Lamb et al. (2010) considered a sample of eight isolated OB stars
(see Table\,\ref{tab:lam} for the list of these stars) in the
Small Magellanic Cloud (SMC). To clarify the origin of these
stars, Lamb et al. (2010) applied almost the same approach as de
Wit et al. (2004, 2005), i.e. they searched for the presence of
unrecognised clusters around isolated massive stars (using the
{\it Hubble Space Telescope} imaging data) and searched for
runaways among them through radial velocity
measurements\footnote{The large distance to the Magellanic Clouds
makes proper motion measurements difficult (see, however, de Mink
et al. 2012), so that detection of high peculiar radial velocities
remains the main tool for revealing runaways in these galaxies.}.

Using the density enhancement and the `friends-of-friends'
algorithms (Davis et al. 1985), Lamb et al. (2010) detected sparse
(approximately parsec scale) concentrations of low-mass stars
around three target stars (AzV\,67, AzV\,106 and AzV\,302), of
which two detections (around AzV\,67 and AzV\,106) are marginal
(see fig.\,2 in Lamb et al. 2010).

Of the remaining five stars, two stars, [MLD95] SMC\,16 and
AzV\,223, were identified as runaways because their radial
velocities exceed by $>30 \, \kms$ the SMC's systemic velocity of
$155 \, \kms$ (see column 3 in Table\,\ref{tab:lam}). This left
Lamb et al. (2010) with three stars, AzV\,58, AzV\,186 and
AzV\,226, apparently formed in complete isolation. Assuming an
isotropic distribution of runaway velocities, Lamb et al.
concluded that these stars could be transverse runaways (i.e.
runaways moving almost in the plane of the sky), but rejected this
possibility because two of them were found to be located within
\hii regions in the line of sight (i.e. may still be in the
regions of their formation). Recall that the same argument was
used by de Wit et al. (2005) to define their four ``best examples
for isolated Galactic high-mass star formation".

Based on the detection of sparse concentrations around three stars
and on the apparently isolated formation of three other stars,
Lamb et al. (2010) concluded that there is no physical $m_{\rm
max}-M_{\rm cl}$ relation, the IGIMF can therefore not be
different from the initial mass function on the scale of a whole
galaxy and only core collapse models of massive star formation are
able to explain these stars, but not the competitive accretion
model. Below we show that these conclusions are {unwarranted and
that other much more likely possibilities to explain the origin of
these stars do exist.

\subsection{Narrowing down the Lamb et al. sample}
\label{sec:nar2}

First, we searched for alternative radial velocity measurements
for all eight stars from the Lamb et al. sample using the VizieR
data base\footnote{http://webviz.u-strasbg.fr/viz-bin/VizieR}. We
found measurements for six stars, three of which (see column 4 in
Table\,\ref{tab:lam}) significantly differ from those reported by
Lamb et al. (2010). This discrepancy could be interpreted as the
indication that three stars from the Lamb et al. sample are
binaries. On the other hand, it suggests that some of them might
be runaways. Further spectroscopic monitoring of these stars is
necessary to judge which of the two possibilities is correct. Note
that the radial velocity measurement for AzV\,223 by Massey et al.
(2009) supports the runaway status of this star. Note also that
AzV\,226 is a fast-rotating star ($v\sin i \approx 300 \, \kms$)
so that the measurement of its radial velocity is less certain
(Evans et al. 2006). Below we show, however, that the runaway
status of just this star can be proven independently.

Then, we searched for bow shocks around all stars from the Lamb et
al. sample using the 24\,$\mu$m mosaic of the SMC obtained with
the {\it Spitzer Space Telescope} within the framework of the {\it
Spitzer} Survey of the Small Magellanic Cloud (S$^3$MC; Bolatto et
al. 2007). We found a bow shock around AzV\,226 (see
Fig.\,\ref{fig:azv}) and thereby prove its runaway status, which
has already been suggested by the radial velocity measurement for
this star (see column 4 in Table\,\ref{tab:lam}). This makes
AzV\,226 the third star in the Magellanic Clouds (besides of
BI\,237 and AzV\,471; see Gvaramadze et al. 2010b, 2011a) whose
runaway status was established through radial velocity
measurements and detections of bow shocks. Recall that the
non-detection of bow shocks around other stars from the Lamb et
al. sample does not exclude the possibility that they are runaways
as well (see Section\,\ref{sec:nar}).

\begin{figure}
\includegraphics[width=8.5cm]{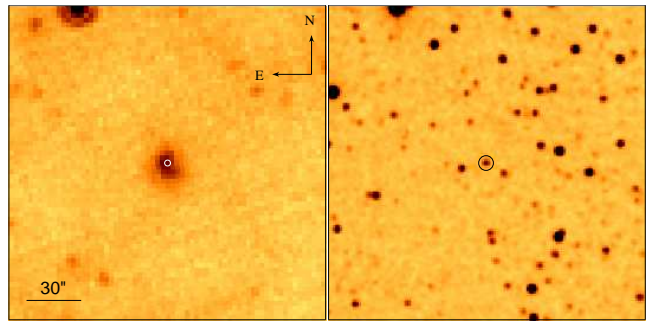}
\centering \caption{{\it Left}: {\it Spizer} $24\,\mu$m image of
the bow shock associated with the O7\,III((f)) star AzV\,226. The
position of AzV\,226 is marked by a circle. {\it Right}: 2MASS J
band image of the same field. At the distance of the SMC, 30
arcsec correspond to $\simeq 8.6$ pc.} \label{fig:azv}
\end{figure}
\begin{figure}
\includegraphics[width=8.5cm]{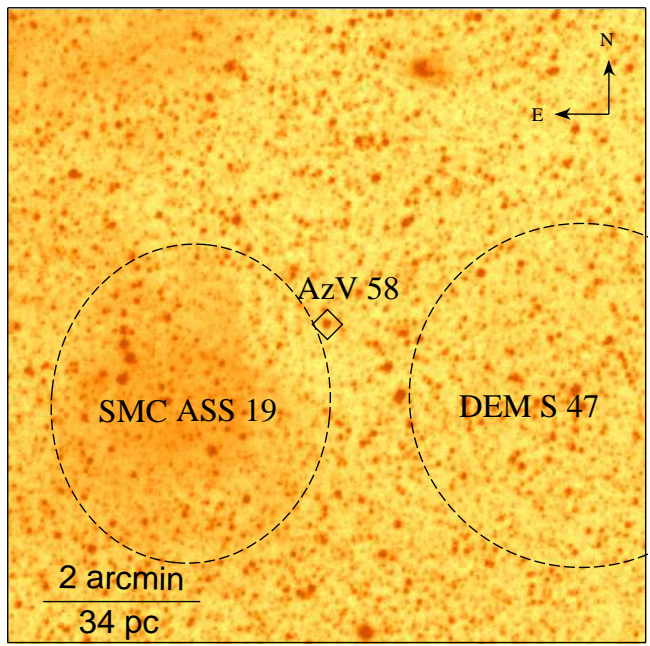}
\centering \caption{DSS-II red band image of the environment of
AzV\,58 (indicated by a diamond). The approximate boundaries of
two nearby associations are shown by dashed ellipse and circle.
See text for details.} \label{fig:dem}
\end{figure}

We found also that the B0.5\,III star AzV\,58 is located at the
north-west periphery of the association SMC\,ASS\,19\footnote{The
approximate boundaries of this and other associations shown in
Figs.\,\ref{fig:dem}-\ref{fig:selier} were taken from the census
of star clusters in the SMC  by Bica \& Dutra (2000).} (see
Fig.\,\ref{fig:dem}). The age of this association of $\approx 8$
Myr (Chiosi et al. 2006) is comparable to that of AzV\,58, which
strongly suggests that SMC\,ASS\,19 is the parent association of
the star. One cannot also exclude the possibility that AzV\,58 was
ejected from the nearby $\approx 8$ Myr old association
DEM\,S\,47.

However, there still remain two stars, AzV\,106 and AzV\,302,
whose radial velocities are comparable to the systemic velocity of
the SMC, and which do not produce (visible) bow shocks. To these
stars one also should add [MLD95] SMC\,16, AzV\,67 and AzV\,186,
whose runaway status remains unclear. One can envisage two
possibilities to explain the origin of these stars.

First, they could either be (transverse) low-velocity or classical
runaways. In the first case, the parent clusters or associations
should be nearby. Using the SIMBAD data
base\footnote{http://simbad.u-strasbg.fr/simbad/}, we found that
the B1\,II (Lamb et al. 2010) star AzV\,106 is located $\approx
4\farcm6$ (or $\approx 80$ pc in projection) from the association
[B91]\,9 (see Fig.\,\ref{fig:azv106}). The age of this association
of $\approx 8$ Myr (Chiosi et al. 2006) is comparable to the age
of the star. If AzV\,106 is a former member of [B91]\,9, then its
peculiar transverse velocity is $\ga 10 \, \kms$, so that this
star is a low-velocity runaway. We found also that the O8.5V (Lamb
et al. 2010) star AzV\,302 is located not far ($\approx 60$ pc in
projection) from the association SMC\,DEM\,118 (see
Fig.\,\ref{fig:azv302}), whose age of $\approx 8$ Myr (Chiosi et
al. 2006) is twice the age of the star. AzV\,302 therefore could
be either a low-velocity rejuvenated star escaping from
SMC\,DEM\,118 (cf. Section\,\ref{sec:nar}) or an ordinary
(transverse) runaway ejected from a more distant stellar system.
Similarly, we found that AzV\,67 is located at $\approx 90$ pc in
projection from the $\approx 10$ Myr old (Chiosi et al. 2006)
cluster [H86]\,119, while [MLD95]\,SMC\,16 and AzV\,186 are
located, respectively, at only $\approx 54$ and 24 pc in
projection from the centres of the $\approx 6$ Myr old (Chiosi et
al. 2006) associations [B91]\,18 and [BS95]\,83. Like AzV\,302,
these three stars could either be low-velocity rejuvenated
runaways (if their possible binary status will be confirmed by
follow-up observations) or ordinary runaways (if they originate in
more distant stellar systems). For example, AzV\,186 might have
been ejected from the $\approx 5$ Myr old (Chiosi et al. 2006)
cluster NGC\,330, which is located at $\approx 110$ pc in
projection from the star. In this case, the peculiar transverse
velocity of AzV\,186 should be $\ga 20 \, \kms$.

Secondly, it is also quite possible that the parent clusters of
some of the five stars already dissolved, especially if these
stars were formed in low-mass clusters with only a few or one
massive star. Such clusters expel their gas rapidly, which results
in the quick dispersal of the systems (see e.g. Kroupa \& Boily
2002; Weidner et al. 2007; Baumgardt \& Kroupa 2007). After the
gas expulsion $\approx 90$ per cent of the stars of the clusters
would be expected to be distributed around the early type stars at
distances of 10 to 50 pc (Weidner et al. 2011). Additionally, a
far more extended population which could have travelled up to 1
kpc should have formed through dynamical few-body interactions in
the clusters. And even if the clusters are still embedded in their
natal gas clouds they are expected to loose up to 20 per cent of
their stars due to dynamical interactions within 5 Myr and more if
the objects are older (Weidner et al. 2011).

\begin{figure}
\includegraphics[width=8.5cm]{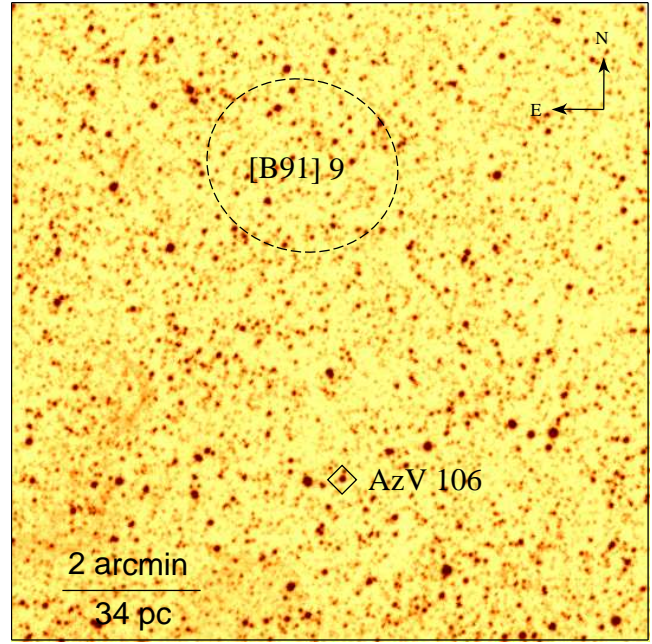}
\centering \caption{DSS-II red band image of AzV\,106 (indicated
by a diamond) and the association [B91]\,9 (indicated by a dashed
circle). See text for details.} \label{fig:azv106}
\end{figure}
\begin{figure}
\includegraphics[width=8.5cm]{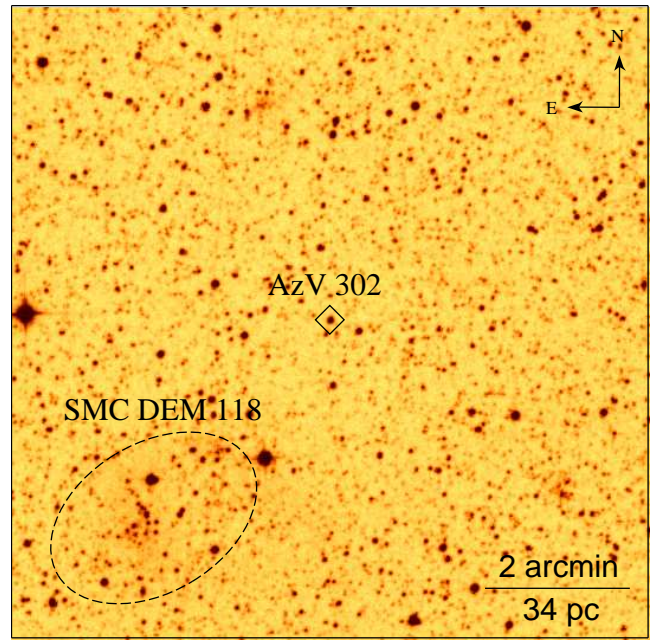}
\centering \caption{DSS-II red band image of the environment of
AzV\,302 (indicated by a diamond), with the position of the
association SMC\,DEM\,118 indicated by a dashed ellipse. See text
for details.} \label{fig:azv302}
\end{figure}

\subsection{Two other candidates for isolated massive star formation}
\label{sec:other}

Recently, Selier et al. (2011) discussed one more candidate for
isolated massive star formation in the SMC, namely, a O6.5-O7\,V
star powering the compact ($\approx 2.2$ pc in diameter) \hii
region LHA 115-N33. Like de Wit et al. (2004), Selier et al.
(2011) searched for a possible parent cluster to this star using
optical images of a rather wide ($\approx 90\times 90$ pc) field
centred on the \hii region. They did not find any statistically
significant stellar cluster around the star, of size larger than 3
pc, which led them to believe that this star ``represents an
interesting case of isolated massive-star formation in the SMC".

\begin{figure}
\includegraphics[width=8.5cm]{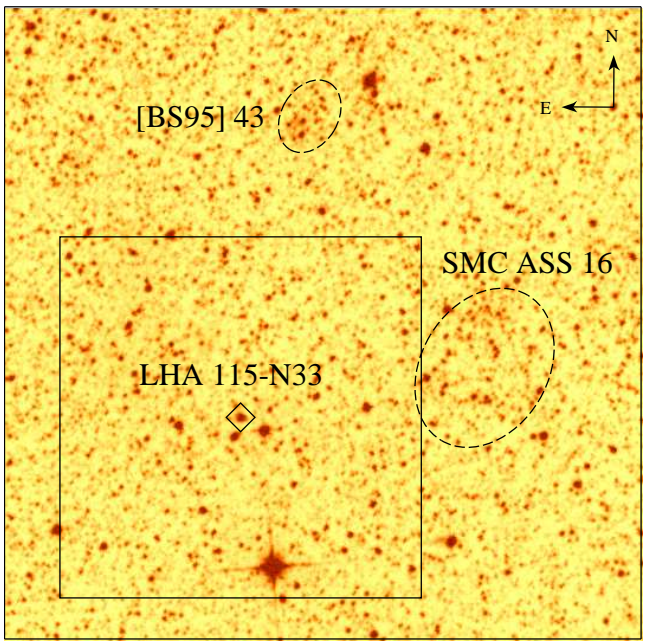}
\begin{center}
   \caption{DSS-II red band image of the field around the \hii region LHA\,115-N33
   (marked by a diamond). The $\approx 90\times 90$ pc field centred on
   the \hii region is shown by a square. The approximate boundaries of two possible
   birth associations to the star powering the \hii region are
   shown by a dashed circle and an ellipse. See text for details.
   }
   \label{fig:selier}
\end{center}
\end{figure}

However, using the SIMBAD data base we found two associations,
SMC\,ASS\,16 and [BS95]\,43, located just outside the field
examined by Selier et al. (2011) at $\approx 3\farcm6$ and
$4\farcm5$  (or $\approx 61$ and 77 pc in projection),
respectively, from the \hii region (see Fig.\,\ref{fig:selier}).
Assuming that the star powering LHA 115-N33 was ejected from one
of these young ($\approx$ 4 and 6 Myr, respectively; Chiosi et al.
2006) associations, one finds that its peculiar transverse
velocity should be $\ga 12-15 \, \kms$, which is a reasonable
velocity.

\begin{figure*}
\includegraphics[width=17.5cm]{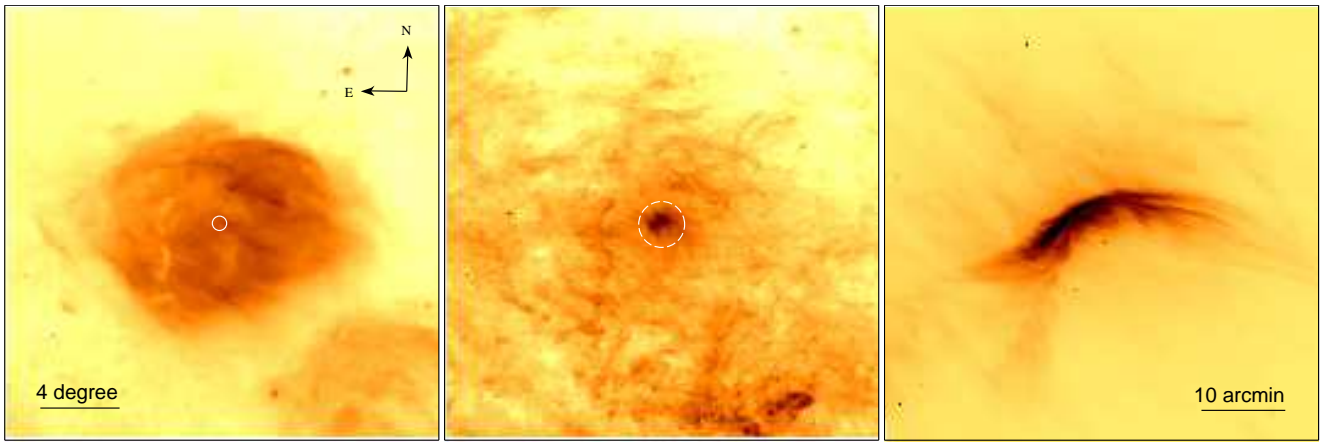}
\begin{center}
\caption{{\it Left}: H$\alpha$ image of the H\,{\sc ii} region
Sh\,2-27 taken from the Southern Hemispheric H$\alpha$ Sky Survey
Atlas (SHASSA; Gaustad et al. 2001). The position of the ionising
O9\,V(e) star $\zeta$ Oph is indicated by a circle. {\it Middle}:
{\it IRAS} 60 $\mu$m image of the same field with the bow shock
generated by $\zeta$ Oph indicated by a dashed circle. The images
were generated by the NASA's SkyView facility (McGlynn, et al.
1998). {\it Right}: {\it Spitzer Space Telescope} $24\,\mu$m image
(Program Id.: 30088, PI: A.~Noriega-Crespo) of the bow shock
around $\zeta$ Oph. The orientation of the images is the same.}
\label{fig:Sh27}
\end{center}
\end{figure*}

In principle, Selier et al. (2011) do not exclude the possibility
that the central star of the \hii region LHA 115-N33 is a runaway.
But as an argument against this possibility they offer the
following reason: ``it seems impossible that a massive star
carries its \hii region during the ejection". On the other hand,
they do not exclude the possibility that a runaway star was
ejected into a molecular cloud, but noted that this situation
``has never been encountered". These two statements are incorrect.

In fact, it is known that any source of ionizing emission moving
through the interstellar medium produces around itself a zone of
ionized gas (i.e. an \hii region) and that in the case of the
supersonic motion the radius of the \hii region is equal to the
Str\"{o}mgren radius (e.g. Tenorio Tagle, Yorke \& Bodenheimer
1979). Whether or not such \hii region would be observable depends
on the number density of the ambient medium and on the total
ionizing-photon luminosity of the star. In this connection, one
can refer to fig.\,5 in Gvaramadze \& Bomans (2008b), which shows
the bow-shock-producing star HD\,165319\footnote{Recall that this
O9.5\,Iab star is considered by de Wit et al. (2005) as one of
``the best examples for isolated Galactic high-mass star
formation".} ejected from the star cluster NGC\,6611 and currently
powering the \hii region RCW\,158 (located $\approx 100$ pc in
projection from NGC\,6611). Another good example of a massive
bow-shock-producing star encountering dense material on its way
through the field and powering an \hii region is the well-known
runaway star $\zeta$\,Oph (Blaauw 1961; Noriega-Crespo et al.
1997; Hoogerwerf, de Bruijne \& Zeeuw 2001) associated with the
\hii region Sh\,2-27 (see Fig.\,\ref{fig:Sh27}).

Therefore it is highly likely that the massive star associated
with LHA 115-N33 is a runaway ejected from one of the two nearby
associations, which met a region of enhanced density (a cloud) on
its way. The high number density of the cloud ($380 \, {\rm
cm}^{-3}$; see Table\,2 of Selier et al. 2011) implies that the
bow shock generated by the star would be too compact to be
resolved even with the {\it Spitzer Space Telescope}, which
provides the best angular resolution ($6''$ at 24\,$\mu$m or
$\approx 1.7$ pc at the distance to the SMC of 60 kpc) among the
modern infrared space telescopes. Indeed, adopting the stellar
mass-loss rate and the wind velocity typical of a O6.5-O7\,V star
of $\approx 10^{-7} \, \msun \, {\rm yr}^{-1}$ and $\approx 2500
\, \kms$ (Mokiem et al. 2007), one finds that for any peculiar
space velocity of the star the angular size of its bow shock would
be at least two orders of magnitude smaller than the angular
resolution of the {\it Spitzer} images (cf. Gvaramadze et al.
2010b).

More recently, Bestenlehner et al. (2011) reported the discovery
of a very massive ($\sim 150 \, \msun$) WN5h star, VFTS\,682,
located $\approx 30$ pc in projection from the very massive star
cluster R136 powering the giant \hii region 30\,Doradus in the
LMC. Bestenlehner et al. (2011) convincingly showed that, like
several other already known very massive runaways in the LMC (e.g.
Evans et al. 2010; Gvaramadze et al. 2010), VFTS\,682 could be a
runaway, but unexpectedly concluded that the apparent isolation of
this star may ``represent an interesting challenge for dynamical
ejection scenarios and/or massive star formation theory".

It is worthy to note that VFTS\,682 is not unique in its very high
mass and the relatively large separation from R136. The very
massive binary R145 (HD\,269928), whose mass is of the same order
of magnitude as that of VFTS\,682 (Schnurr et al. 2009), is
located $\approx 20$ pc in projection from R136. The large offset
of R145 from R136 could be interpreted as the indication that the
binary was recoiled from the parent cluster due to an energetic
three-body gravitational interaction in the cluster's core and
that a massive runaway star was ejected in the opposite direction
(Gvaramadze \& Gualandris 2011; cf. Fujii \& Portegies Zwart
2011). Interestingly, such a star indeed exists just on the
opposite side of 30\,Doradus (see fig.\,12 in Gvaramadze \&
Gualandris 2011). This star, Sk$-69\degr$206, located $\approx
240$ pc to the west of R136, was identified as a runaway via
detection of its associated bow shock, whose orientation is
consistent with the possibility that Sk$-69\degr$206 was ejected
from 30\,Doradus (Gvaramadze et al. 2010b). Moreover, if one
assumes that Sk$-69\degr$206 and R145 were ejected from R136 owing
to the same three-body encounter, then the conservation of the
linear momentum implies that the mass of Sk$-69\degr$206 should be
$\approx 10-15 \, \msun$, which is consistent with the approximate
spectral type of this star of B2 (Rousseau et al. 1978).

It is therefore likely that VFTS\,682 is a former massive binary
which was recoiled from R136 in the course of a strong dynamical
three-body encounter in the dense core of R136 and merged into a
single star (e.g. because of encounter hardening). The same
conclusion also follows from high-precision $N$-body simulations
of R136-like initially fully mass-segregated and binary-rich
clusters (Banerjee et al. 2012), which show conclusively that
dynamical ejections of very massive stars with kinematic
properties similar to those of VFTS\,682 are common, and that
these very massive runaways represent the most probable type of
runaways produced by such clusters.


\section{Summary and conclusion}
\label{sec:sum}

In this paper, we examined claims of the existence of isolated
massive star formation in the Milky Way and the Magellanic Clouds.
These claims are often used to support the {\it in situ} proposal
on the origin of massive stars. Our goal was to check whether the
best candidates for isolated formation of massive stars are
actually runaway stars, and therefore were formed in embedded
clusters and subsequently ejected into the field because of
dynamical few-body interactions or binary-supernova explosions.

Several indicators can be used to reveal the runaway status of the
field O stars, namely, the high (say, $>30 \, \kms$) peculiar
transverse and/or radial velocity of the stars or the presence of
bow shocks around them. Detection of high radial velocities and/or
bow shocks is especially useful for revealing the runaway status
of distant stars, whose proper motion measurements are still not
available (e.g. in the Magellanic Clouds) or are measured with a
low significance. For Galactic candidates for isolated massive
star formation (which are relatively nearby objects) the new
reduction of the {\it Hipparcos} data can also be used to search
for their high transverse peculiar velocities.

Careful examination of the existing observational data showed that
all but one of the best Galactic candidates for isolated massive
star formation are in fact high-velocity and/or
bow-shock-producing (i.e. runaway) stars. The only star,
HD\,124314, for which we derived a low peculiar (transverse)
velocity and did not detect a bow shock\footnote{The non-detection
of bow shocks around field stars does not exclude their runaway
status, but could be caused by the motion of these stars through
low-density, hot interstellar gas, which makes the bow shocks
unobservable or even precludes their formation at all.} is a
candidate single-line spectroscopic binary. Thus, it is very
likely that HD\,124314 is a post-supernova binary system, which
was dynamically ejected from the parent cluster and whose space
velocity was reduced (and re-oriented) due to the kicks caused by
the mass loss from the system and the asymmetry of the supernova
explosion. Moreover, the mass transfer in the binary system prior
to the supernova explosion might have significantly rejuvenated
HD\,124314, so that the actual distance travelled by this star
could be much larger than that inferred from the apparent age of
the star and its peculiar transverse velocity. Thus, HD\,124314
might belong to a population of O stars (the descendants of
runaway massive binaries) that {\it must} exist in the field and
whose low space velocities and/or young ages are in apparent
contradiction with the large separation of these stars from their
parent clusters and/or the ages of these clusters, and which can
be mistakenly considered as having formed {\it in situ}.

We also found that the candidates for isolated massive star
formation in the Magellanic Clouds either possess high peculiar
radial velocities (and therefore are runaways; one of these stars
is associated with a bow shock as well) or are located not far
from young stellar associations (and therefore might be runaways
moving almost in the plane of the sky). One of the latter stars,
AzV\,302, is about two times younger than the nearby association.
This star could either be a rejuvenated low-velocity runaway or a
high-velocity (transverse) runaway ejected into the field from a
more distant stellar system. It is also possible that the parent
cluster of AzV 302 (as well as the birth clusters of some other
field O stars) already dissolved, especially if this star was
formed in a low-mass cluster with only a few or one massive star.
Such clusters expel their gas rapidly, which results in the quick
dispersal of the systems. We argue also that some field O stars
could be detected in optical wavelengths only because they are
runaways, while their cousins residing in the deeply embedded
parent clusters might still remain totally obscured.

The main conclusion of our study is that there is no significant
evidence for massive stars formed in isolation. While it can never
be proven to absolute certainty that a particular massive star was
formed in a star cluster, the sum of the evidence, and in
particular the known and well understood stellar dynamical
processes, does not support isolated massive star formation as
occurring.


\section{Acknowledgements}
We are grateful to the referee for comments that allowed us to
improve the content and the presentation of the paper. V.V.G.
acknowledges financial support from the Deutsche
Forschungsgemeinschaft. C.W. acknowledges financial support
through the CONSTELLATION European Commission Marie Curie Research
Training Network (MRTN-CT-2006-035890).


\end{document}